\documentclass[12pt]{article}

\usepackage{booktabs}
\usepackage[english]{babel}
\usepackage{amsmath,amssymb,amsbsy,amstext, amsthm, simplewick}
\usepackage{graphicx}
\usepackage{amsfonts}
\usepackage{amssymb}
\usepackage{upgreek}
\usepackage{simplewick}
\usepackage{exscale,relsize}
\usepackage{bbding}
\usepackage{cite}

\newcommand{\Expect}[1]{\left\langle #1 \right\rangle}

\newcommand{\bea}{\begin{eqnarray}}
\newcommand{\eea}{\end{eqnarray}}

\newcommand{\mpl}{m_{\mbox{\tiny Pl}}}

\newcommand{\beq}{\begin{equation}}
\newcommand{\eeq}{\end{equation}}
\newcommand{\be}{\begin{equation}}
\newcommand{\ee}{\end{equation}}

\newcommand{\ra}{\rightarrow}
\newcommand{\lp}{\left(}
\newcommand{\rp}{\right)}

\newcommand{\nn}{\nonumber}

\def\gev{\, {\rm GeV}}

\def\ev{\, {\rm eV}}

\def\ev{\, {\rm eV}}

\begin{document}

\thispagestyle{empty}

\vspace*{1.31cm}

\begin{center}
{\LARGE \bf {Constraints on bosonic dark matter from   }}\\
\vspace*{0.3cm}
{\LARGE \bf {observations of old neutron stars}}\\

\vspace*{1.61cm} {\large
Joseph Bramante$^{\bigstar,}$\footnote{\tt
bramante@hawaii.edu},
Keita Fukushima$^{\bigstar,}$\footnote{\tt keitaf@hawaii.edu},and
Jason Kumar$^{\bigstar,}$\footnote{\tt jkumar@hawaii.edu}}\\
\vspace{.5cm}
{ $^{\bigstar}$ Department of Physics and Astronomy,
University of Hawaii, 2505 Correa Road, Honolulu, HI, 96822, USA}\\

\vspace{1.5cm}

{\Large ABSTRACT}
\end{center}

{
Baryon interactions with bosonic dark matter are constrained by the potential for dark matter-rich neutron stars to collapse
into black holes.
We consider the effect of dark matter self-interactions and dark matter annihilation on these bounds, and treat the evolution
of the black hole after formation.
We show that, for non-annihilating dark matter, these bounds extend up to
$m_X \sim 10^{5-7}$ GeV, depending on the strength of self-interactions.
However, these bounds are completely unconstraining for annihilating bosonic dark matter
with an annihilation cross-section of
$\langle \sigma_a v \rangle \gtrsim 10^{-38}~{\rm cm^3 /s }$.  Dark matter decay does not significantly affect these bounds.
We thus show that bosonic dark matter accessible to near-future direct detection experiments must participate in an annihilation or
self-interaction process to avoid black hole collapse constraints from very old neutron stars.
}
\vspace{4cm}
\vfill

\setcounter{page}{0} \setcounter{footnote}{0}


\section{Introduction}

Dark matter (DM) has been detected only via gravitational interactions. While there is overwhelming cosmological and
astronomical evidence for new matter which could have a weak coupling to standard model fermions, the mass and couplings
of this dark matter are not yet established.
Asymmetric dark matter (ADM) has been proposed as a compelling framework
to explain both the dark matter abundance and the baryon
asymmetry~\cite{Nussinov:1985xr,Hooper:2004dc,Gudnason:2006ug,Khlopov:2007ic,Foadi:2008qv,Kaplan:2009ag,Cohen:2010kn,Shelton:2010ta,Falkowski:2011xh,
Buckley:2010ui,Davoudiasl:2010am,Blennow:2010qp,Hall:2010jx,Haba:2010bm,
Frandsen:2011kt,Graesser:2011wi,Cheung:2011if,MarchRussell:2011fi,
Dutta:2010va,Gu:2010ft,Allahverdi:2010rh,DelNobile:2011je,
Tulin:2012re,Iminniyaz:2011yp,Buckley:2011ye,Buckley:2011kk,
Cui:2011qe,Davoudiasl:2011fj,Lin:2011gj,Kumar:2011np,Kaplan:2011yj,Cirelli:2011ac,MarchRussell:2012hi}.


It has been pointed out that models of asymmetric dark matter can be tightly constrained by the existence of
old neutron stars~\cite{deLavallaz:2010wp,kouvaris,kouvaris2,mcdermott,bertone,mccullough,Press:1985ug,goldman}.
The basic point is
that dark matter will be captured by neutron stars due to DM-neutron scattering;
if dark matter does not annihilate or decay, then it will continue to accumulate until it collapses into a black hole.
The observation of old neutron stars can thus bound the dark matter-neutron scattering cross section ($\sigma_{nX}$) for
models with no dark matter annihilation or decay.
It has also
been demonstrated that bounds on non-annihilating dark matter shift with the introduction of self-interactions
terms~\cite{Spergel:1999mh,Randall:2007ph,Slepian:2011ev,Kouvaris:2011gb,Guver:2012ba}.
In this work, we will study a more general question.  We will focus on the range of $\sigma_{nX}$, dark matter
annihilation cross section ($\sigma_{a}v$), dark matter decay rate ($\Gamma$), and self-interaction strength which
could be consistent with observations of old neutron stars.

As an initial point, we note that asymmetric dark matter is not necessarily non-annihilating.  Asymmetric dark matter
requires that the dark matter particle can be distinguished from the anti-particle, which implies the
existence of a continuous unbroken symmetry under which the dark matter is charged.  It is this symmetry which forbids the
fermion Majorana mass term.  However, asymmetric dark matter need not be the lightest particle charged
under this continuous symmetry.  For dark matter to be stable, it must be the lightest particle charged under some symmetry, but this
may be a distinct $Z_2$ symmetry.  In this case, self-annihilation of the asymmetric dark matter would not be forbidden.

Here we summarize the important features of the following analysis:
\begin{itemize}
\item[I.] {\it Dark matter accumulation.}  The rate at which a neutron star captures dark matter ($C_X$)
depends on the local dark matter density ($\rho_X$),
the dark matter-neutron scattering cross section ($\sigma_{nX}$), and the dark matter mass ($m_X$).
The number of dark matter particles in a neutron star can be depleted by dark matter decay or annihilation.
\item[II.] {\it Black hole formation.}  Bosonic dark matter collected in a neutron star will form a black hole
if the total energy is minimized at a
radius which is less than the Schwarzschild radius.
This condition is usually fulfilled by requiring that the dark matter be self-gravitating
and exceed the Chandrasekhar limit.  The Chandrasekhar limit grows with the strength of repulsive
self-interactions.
\begin{itemize}
\item[a.] If the dark matter cannot thermalize quickly enough, it will not form a black hole within the lifetime of the neutron star
(see Appendix~\ref{thermapp}).
\item[b.] Thermalized dark matter will collect within a radius $r_{th}$, which determines the number of
dark matter particles needed for the dark matter to self-gravitate.  But if dark matter forms a Bose-Einstein condensate (BEC),
then it will collect within a much smaller radius ($r_c$) and fewer particles will be needed before the dark matter
becomes self-gravitating and gravitational collapse occurs.
\end{itemize}
\item[III] {\it Destruction of the neutron star.}  The neutron star will be destroyed if the black hole grows large enough to
consume the neutron star.  The black hole will accrete baryonic and dark matter, but will emit Hawking radiation.  To destroy
the neutron star, the black hole must be large when it forms.
\begin{itemize}
\item[a.] If dark matter forms a BEC, then dark matter captured by the neutron star after the formation of a black hole will be
efficiently accreted by the black hole, potentially compensating for the effect of Hawking radiation.
\item[b.] If dark matter has a repulsive self-interaction, then the number of particles needed to form a black hole will increase.
The resulting black hole, when formed, may then be large enough to continue growing.
\end{itemize}
\end{itemize}

In section 2, we describe the accumulation of dark matter in neutron stars, including effects
from dark matter decay and annihilation.  In section 3, we describe the formation of a black hole
from dark matter in neutron stars, including the effects of self-interactions.  In section
4, we describe evolution of a black hole which has formed in a neutron star, including the
effects of baryonic and dark matter accretion and Hawking radiation.  In section 5 we
find the constraints on the parameter space of bosonic dark matter models from observations
of old neutron stars.  We conclude with a discussion of our results in section 6.

\section{Dark Matter Accumulation in a Neutron Star}

The dark matter capture rate ($C_X$) of neutron stars is given in \cite{Gould:1987ir,mcdermott}.
For $m_X < {\cal O}(10^2)~\ev$ (assuming $T=10^5~{\rm K}$), dark matter within
an old neutron star will be depleted by evaporation, and the rate of dark matter accumulation will also be suppressed.
The probability for a dark matter particle to
scatter while passing through a neutron star is given by~\cite{Kouvaris:2007ay}
\beq
P =  1- \exp \left[-\int \eta_{n}\sigma_{nX} dl \right]
\eeq
where $\eta_n$ is the neutron number density and
the integral is taken over the path of the dark matter particle through the neutron star.
For small $\sigma_{nX}$, we have $P \sim \eta_n \Delta l \sigma_{nX}$.  But $P \rightarrow 1$ for
$\int \eta_{n}\sigma_{nX} dl \gg 1$.  Taking a neutron star to have radius $R=10.6~{\rm km}$ and baryonic density
$\rho_b \sim 7.8 \times 10^{38} ~\rm{GeV}/cm^3$~\cite{Clark:2002db,deLavallaz:2010wp,mcdermott}, this saturation occurs
for $\sigma_{nX} > \sigma_{sat} \sim {2.1 \times 10^{-45}~\rm{cm^2}} $.

For $m_X >  \rm{GeV}$ the capture rate is~\cite{mcdermott}
\beq
C_X \sim 2.3 \times 10^{45}~{\rm Gyr}^{-1} ~ \lp \frac{\rm{GeV}}{m_X} \rp  \lp \frac{\rho_X}{{\rm 10^3~ GeV/cm^3}} \rp
f(\sigma_{nX}) \beta(m_X,m_N,v_{esc},\bar{v}), \label{cxlarge}
\eeq
while for $m_X < ~ \rm{GeV}$, due to the effects of Pauli blocking, the capture rate is~\cite{mcdermott},
\beq
C_X \sim 3.4 \times 10^{45}~{\rm Gyr}^{-1}  \lp \frac{\rho_X}{{\rm 10^3 ~GeV/cm^3}} \rp  f(\sigma_{n_X}) \beta(m_X,m_N,v_{esc},\bar{v}),
\eeq
where $\rho_X$ is the ambient density of dark matter.  The factor $f(\sigma_{nX})$ is given by
$f = \sigma_{nX} / \sigma_{sat}$  for $\sigma_{nX} \leq \sigma_{sat}$,
and $f=1$ for $\sigma_{nX} > \sigma_{sat}$. The factor
\bea
\beta(m_X,m_N,v_{esc},\bar{v})
&=&  1- \frac{1- \exp \left[ -6 (v_{esc}^2  /\bar{v}^2) (\mu / (\mu -1)^2) \right]}
{6 (v_{esc}^2  /\bar{v}^2) (\mu / (\mu -1)^2)}
\eea
will take the value $\beta \sim 1$ for dark matter masses $m_X \lesssim 10^6 ~ {\rm GeV}$ and for typical neutron star parameters,
where $\mu \equiv m_X/m_N$, $v_{esc} \simeq 1.8 \times 10^5 ~{\rm km/s}$ is the escape velocity from the surface of the
neutron star, and $\bar{v} \sim 220 ~{\rm km/s}$ is the ambient dark matter average velocity \cite{deLavallaz:2010wp,mcdermott,Gould}.

If $t_{ns}$ is the lifetime of a neutron star, then the number of particles accumulated over that lifetime,
$N_{acc}(t_{ns})$, is determined by the dark matter capture rate, decay rate, and the rate of dark matter
annihilation.
We will first consider the case where dark matter does not annihilate, but does decay at rate $\Gamma = \tau^{-1}$.
In this case, the number of accumulated dark matter particles can be written as
\bea
N_{acc}^{(decay)} = C_X \tau \lp 1- e^{- t_{ns} / \tau}  \rp. \label{cxtau}
\eea
The tightest model-independent constraints on dark matter decay come from an analysis of the stability of dark matter halos. Measurements
of halo mass-concentration and galaxy-cluster mass compared with simulations of dark matter halo mass distributions disturbed by dark matter
decay constrain any dark matter lifetime to $\tau > 10 {\rm ~ Gyr} $ for all velocities of the dark matter decay products~\cite{Peter:2010jy}.
Constraints on dark matter decay that heats the CMB are tighter for decay products with $v \gtrsim 0.6 \rm{c}$~\cite{Aoyama:2011ba}.
We see that for $\tau \sim 10 ~{\rm Gyr}$ and a neutron star lifetime of $t_{ns} \sim 10 ~{\rm Gyr}$, the number of
accumulated dark matter particles is only suppressed by an ${\cal O}(1)$ factor.
Because the minimum allowed dark matter lifetime is on the order of the lifetime of a neutron star, dark matter decay
does not significantly alter the amount of accumulated dark matter, and thus does not significantly alter constraints
arising from observations of neutron stars.

Henceforth, we will assume that dark matter does not decay.
The accumulated number of dark matter particles can then be approximated as~\cite{Griest:1986yu,Jungman:1995df}
\bea
\frac{d N_{acc}}{dt} &\approx& C_X - \frac{\langle \sigma_a v \rangle N_{acc}^2}{V_{th}} \nn \\
\ra N_{acc} &\approx& \sqrt{\frac{C_X V_{th}}{\langle \sigma_a v \rangle} }
{\rm Tanh} \left[ \sqrt{\frac{C_X \langle \sigma_a v \rangle}{V_{th}}} t_{ns} \right], \label{nacc}
\eea
where $V_{th} = (4/3) \pi r_{th}^3$ is the volume within which the dark matter is thermalized, here assumed to be of constant density, and
$\langle \sigma_a v \rangle $ is the annihilation cross section.
If the effect of self-interactions is small, the thermalization radius $r_{th}$ can be written as~\cite{kouvaris,mcdermott,goldman},
\beq
r_{th} = 240~ {\rm{cm}} \lp \frac{T}{10^5 ~\rm{K}} \cdot \frac{\rm{GeV}}{m_X} \rp^{1/2}.
\eeq
It is useful to determine the range of physical parameters for which the argument of the hyperbolic tangent in eq.~\ref{nacc} is greater than unity,
indicating that the collection and annihilation of dark matter in the
neutron star has reached an equilibrium. For a more complete treatment, see Appendix~\ref{equilib}.

\section{Black Hole Formation} \label{bhf}
In order to form a black hole, the dark matter collected in a neutron star must become dense enough that the energy of the
dark matter is minimized as the radius of the dark matter distribution approaches zero.
If $N_{DM}$ bosonic dark matter particles of mass $m_X$ are confined to a sphere of radius $r$, then
the energy of a boson is approximately given by
\bea
E \sim \frac{1}{r} - \frac{ G m_X^2 N_{DM} }{r} + \frac{2 \pi G \rho_b m_X  r^2}{3}. \label{effpot}
\eea
The first term is the relativistic kinetic energy, and the second and third terms are the
gravitational potential energy due to DM-DM interactions and DM-baryon interactions, respectively.
The requirement of ``self-gravitation" ensures the second
term of eq.~\eqref{effpot} is larger than the third, so the second term will dominate as the dark matter collapses. The
Chandrasekhar limit then corresponds to the requirement that the second term dominate the kinetic term.

However, this effective Chandrasekhar limit depends on the total local potential of the dark matter and is modified if
dark matter has self-interactions.  A $\lambda |\phi|^4$ term is generally not forbidden by any symmetry
of the theory (in the absence of higher dimension terms, stability of the potential would require this interaction to
be repulsive, $\lambda \geq 0$).  With this self-interaction~\cite{kouvaris},
the number of self-gravitating particles required to form a black hole will be~\cite{mielke,jetzer}
\bea
N_{chand} = \frac{2 \mpl^2}{\pi m_X^2} \lp 1 + {\lambda \over 32\pi} {\mpl^2 \over m_X^2} \rp^{1/2}. \label{nchand}
\eea
Of course this expression reduces to the simpler limit $N_{chand} \sim \mpl^2/m_X^2$ for non-interacting bosons when $\lambda=0$.
Note, however that if $\lambda / 32\pi > m_X^2 / \mpl^2$, then $N_{chand} \sim \lambda^{1/2} \mpl^3/m_X^3$.  As
the $\lambda |\phi|^4$ interaction is not forbidden by any symmetry, there is no reason to expect $\lambda$ to be very small.
Thus, unless $m_X$ is quite large, one would expect the Chandrasekhar limit to be dominated by the interaction term.  In this
case, the Chandrasekhar limit on the number of particles is suppressed from the fermion case
($N_{chand}^{(ferm.)} \sim \mpl^3/m_X^3$) by a factor $\lambda^{1/2}$.

The dark matter particles will become self-gravitating when their density exceeds that of the baryons
in the neutron star.  If the dark matter is confined to a region of radius $r$, then the number
of dark matter particles required to achieve self-gravitation is given by
\bea
N_{s-g} (r) \simeq \frac{4 \pi r^3}{3 m_X}  \rho_b, \label{nsg}
\eea
where $\rho_b \sim 7.8 \times 10^{38} ~\rm{GeV}/cm^3$ is taken as the baryon density in a neutron star.

If dark matter forms a Bose-Einstein condensate~\cite{kouvaris,mcdermott}, then a large fraction of the dark matter will be confined to a
radius which is much smaller than the thermalization radius of the dark matter.  Thermalized bosonic matter at the core
of a neutron star will form a BEC if the number of thermalized particles exceeds
\beq
N_{BEC} = \zeta \lp \frac{3}{2} \rp \lp \frac{m_X T}{2 \pi} \rp^{3/2} \lp \frac{4 \pi r_{th}^3}{3} \rp \approx 10^{36} \lp \frac{T}{10^5 ~\rm{K}} \rp^3.
\eeq

The effect of self-interactions on the formation of a BEC is still not completely understood.  In principle,
self-interactions can affect the critical temperature
and the size of the BEC state.  A complete study of these effects is beyond the scope of this
work.  Instead, we will assume that the critical temperature and the size of the BEC state are unchanged by
self-interactions of the magnitude which we will consider.  If, due to self-interactions, dark matter does
not form a BEC, then the analysis of Appendix~\ref{AppendixNonBEC} would be relevant.

If the number of dark matter particles in the BEC phase is small, and the gravitational potential energy
is dominated by the baryonic contribution, then the size of the BEC, $r_c$,  can be
approximated by equating the magnitude of the non-relativistic kinetic energy and gravitational potential
energy~\cite{kouvaris,mcdermott}, yielding:
\bea
r_c &=& \lp \frac{3}{8 \pi G m_X^2 \rho_b } \rp^{1/4} = 1.5 \times 10^{-4}~ {\rm cm} \lp \frac{{\rm GeV}}{m_X} \rp^{1/2} .
\eea
We have assumed that the $\lambda \phi^4$ contribution is small.  Considering the ground state to have size $r_c$,
the contribution of the $\lambda \phi^4$ term to the energy of the BEC state scales approximately
as $\propto \lambda N^2 / r_c^3 m_X^2$.  This contribution is negligible for
$\lambda \ll 10^{-18} (m_X /\gev)^3$, implying that the value of the critical temperature and the
size of the BEC state are essentially unchanged.
From eq.~\eqref{nsg} we find that the number
of particles in the BEC phase required for self-gravitation is
\bea N_{s-g}^{(BEC)} =  10^{28}
\lp \frac{\rm{GeV}}{m_X} \rp^{5/2}.
\eea
Assuming $~T=10^5 ~ \rm{K}$, one finds $N_{chand}  > N_{s-g}$ if
$m_X > 4 \times 10^{-21}~ {\rm GeV}$.
For all $m_X$ of interest BECs will become self-gravitating well before they reach the Chandrasekhar limit.

We may write the number of dark matter particles needed for black hole formation in the
BEC phase as $N_{BHforms}^{(BEC)}$.
We find
\bea
N_{BHforms}^{(BEC)}(m_X, \lambda, T) &=&
N_{chand} + N_{BEC} .
\eea
Note that, if dark matter forms a BEC, the neutron star must collect $N_{BEC}$ particles which lie within
$r_{th}$ and cause the formation of the BEC, as well as an additional $N_{chand}$ particles which fall into a BEC
of size $r_c$ and which collapse to form a black hole.
For sufficiently large $m_X$, one finds $N_{s-g}(r_{th}) > N_{BEC}$, in which case
the dark matter collected within the thermalization radius will self-gravitate before enough dark matter
is collected to form a BEC.
This leads to the possibility that dark matter within the thermalized region will collapse without forming a
BEC~\cite{Guver:2012ba,Fan:2012qy}.  However, as dark matter in the thermalized region collapses it will also lose energy,
which can result a lower temperature and higher density.  This may subsequently lead to
the formation of a BEC~\cite{Kouvaris:2012dz}.

\section{Hawking Radiation and Neutron Star Destruction} \label{hrnsd}

The formation of a black hole within an old neutron star is not necessarily in conflict with observation  -- the black hole
must also absorb the neutron star in a time much shorter than the lifetime of the neutron
star, without first evaporating through Hawking radiation.
We build on the analysis of
\cite{kouvaris,mcdermott}, but also consider the effect of dark matter accretion and repulsive self-interactions on the growth of the
black hole.

The evolution of the black hole's mass is governed by the equation
\beq
\frac{dM_{bh}}{dt} = \frac{4 \pi \rho_b (G M_{bh})^2}{v_s^3} + \lp \frac{d M_{bh}}{dt} \rp_{DM} - \frac{1}{15360 \pi (G M_{bh})^2} \label{fullhawk},
\eeq
where $M_{bh}$ is the mass of the black hole and $v_s $ is the sound speed of the neutron star (we take $v_s /c \sim 0.1 $~\cite{mcdermott}).
The first term
on the right hand side of eq.~\eqref{fullhawk} is the Bondi accretion rate for baryonic matter, the second term is the rate
at which the black hole accretes dark matter, and the last term is the
Hawking radiation rate. $\lp dM_{bh}/dt \rp_{DM}$ will depend not only on how quickly the neutron star captures dark matter, but also on how quickly
the black hole absorbs new dark matter captured by the neutron star.  The initial black hole mass, $M_{bhi}$ is given by
$m_X N_{s-g}$ ($N_{s-g}>N_{chand}$) and $m_X N_{chand}$ ($N_{s-g}<N_{chand}$).

\subsection{Rate of Black Hole Growth and Destruction}

If a black hole within a neutron star begins to grow, then the baryonic accretion rate will increase as
$M_{bh}^2$, while the Hawking radiation rate will decrease as $M_{bh}^2$.
In this case, we can approximate the time it will take for the black hole to consume the neutron star by assuming that
the baryonic accretion rate dominates, neglecting dark matter accretion and Hawking radiation.  We then find
\bea
\frac{dt}{dM} &=& \frac{v_s^3}{4 \pi \rho_b  {\lp G M \rp}^2} \nn \\
&\ra & t_{nscollapse} = \frac{v_s^3}{4 \pi \rho_b G^2} \lp \frac{1}{ M_{bhi}} - \frac{1}{ M_{bhi}+M_{ns}} \rp
\sim \frac{v_s^3}{4 \pi \rho_b  { G^2 M_{bhi}}},
\eea
where $M_{ns} \sim 3.3 \times 10^{57} ~\rm{GeV}$ is the mass of a heavy neutron star~\cite{Kiziltan:2010ct}.
We consider the rate of collapse for a black hole
with initial mass given by
\bea
M_{bhi}^{(BEC)} &=& m_X N_{chand} = 9.5 \times 10^{37}~\gev \lp \frac{ \rm GeV}{m_X} \rp \lp 1 + {\lambda \over 32\pi} {\mpl^2 \over m_X^2} \rp^{1/2}.
\label{eq_MBHI_BEC}
\eea
This initial mass yields a neutron star collapse time of
\bea
t_{nscollapse}^{(BEC)} &=& 2.6 \times 10^{5} ~{\rm{years}} \lp \frac{m_X}{\rm GeV} \rp \lp 1 + {\lambda \over 32\pi} {\mpl^2 \over m_X^2} \rp^{-1/2},
\eea
For all relevant regions of parameter space, this time of collapse will be small compared to the lifetime of an old neutron star.

Similarly, if a black hole begins to shrink, the baryonic accretion rate will quickly become small, while the Hawking
radiation rate will grow rapidly.  Considering only the Hawking radiation rate, we find
\bea
{dt \over dM} &=& -15360 \pi (G M)^2 ,
\nonumber\\
t_{evap} &=& 5120 \pi G^2 M_{bhi}^3 .
\eea
We then find
\bea
t_{evap}^{(BEC)} &=& 5120 \pi G^2 \left(9.2 \times 10^{37} {\gev^2 \over m_X} \right)^3
=13 {~\rm Gyr} \left({\gev \over m_X} \right)^3 .
\eea
We thus see that the black hole will evaporate in a time much shorter than the
lifetime of the neutron star if $m_X \gg 1~\gev$, for dark matter which forms a BEC.

\subsection{Black Hole Accretion of Dark Matter in the BEC Phase}
In order for dark matter to efficiently fall into a black hole after being captured by the neutron star, the impact parameter of the black hole must
be small compared to the radius within which the dark matter settles \cite{mcdermott}. In other words, additional dark matter will accumulate in the
black hole at the rate it enters the neutron star if the size of the region where the dark matter particles settle is small compared to the
black hole's impact parameter,
$b_{infall} = 4GM_{bh}/v_\infty$. Here $v_\infty$ is a dark matter particle velocity on approach to the black hole.

After a black hole is formed from a BEC, $N_{BEC}$ dark matter particles will remain within radius $r_{th}$, and any
further dark matter particles that collect in the star will fall into the BEC state.
It can be shown that the radius of the BEC is smaller than the black hole's
impact parameter \cite{mcdermott}, specifically $b_{infall} \sim 4 r_{c}$, so in the case that all incoming dark matter settles into a BEC state, it
will be efficiently captured by a black hole. Thus for dark matter which forms a BEC, the black hole dark matter accretion rate will equal the neutron
star dark matter capture rate,
\bea
\lp \frac{dM_{bh}}{dt} \rp_{DM} \sim C_X m_X.
\eea
The black hole will continue to grow if
\bea
C_X m_X &>& \frac{1}{15360 \pi (G m_X N_{chand})^2} -\frac{4 \pi \rho_b (G m_X N_{chand})^2}{v_s^3}
\nonumber\\
C_X &>&
{1\over {\rm Gyr}}
\left[2.4 \times 10^{36} \lp \frac{ m_X}{{ \rm GeV}}  \rp \left(1+{\lambda \over 32\pi} {\mpl^2 \over m_X^2} \right)^{-1}
\right.
\nonumber\\
&\,& \left.
-1.5\times 10^{42} \left(\gev \over m_X \right)^3
\left(1+{\lambda \over 32\pi} {\mpl^2 \over m_X^2} \right) \right] .
\label{NexcBEC}
\eea
If the black hole begins to grow, it will quickly absorb the entire neutron star.  On the other hand,
for the range of masses in which it is possible
for Hawking radiation to dominate, the black hole will evaporate quickly.

A possible exception to even this bound arises if one assumes that Hawking radiation preferentially heats bosonic dark matter via
dark-sector radiation. We refer to a detailed discussion of this effect in \cite{mcdermott}.  However, we note that, since a growing
black hole will in fact grow rapidly, the Hawking radiation rate will quickly become small, implying that there will be very little
heating of the dark or baryonic matter due to Hawking radiation.

Bosonic dark matter with a large enough repulsive self-interaction
cross section will have a larger mass at Chandrasekhar collapse and can avoid forming a small black hole that evaporates too quickly
to destroy the neutron star.
The result is an interesting phenomenon: very small repulsive self-interactions tighten neutron star collapse constraints on bosonic dark
matter, but larger repulsive self-interactions loosen the same constraints.

\section{Bosonic Dark Matter Bounds From Neutron Star Collapse}

In this section we determine the constraints on $\sigma_{nX}$ for bosonic dark matter arising from the existence of
old neutron stars, including the effects of dark matter self-interactions, dark matter annihilation, and dark matter
accretion onto black holes.
The exclusion contour bounds the region
\bea
N_{acc} (\sigma_{nX}, m_X, \langle \sigma_a v \rangle, \rho_X, t_{ns}, T) &>& N_{BHforms} (m_X, \lambda, T),
\nonumber\\
\left. {dM_{BH} \over dt}\right|_{M_{BH}=M_{BHi}} &>& 0.
\label{thebound}
\eea

\begin{figure}[!ht]
\centering
\includegraphics[scale=.75]{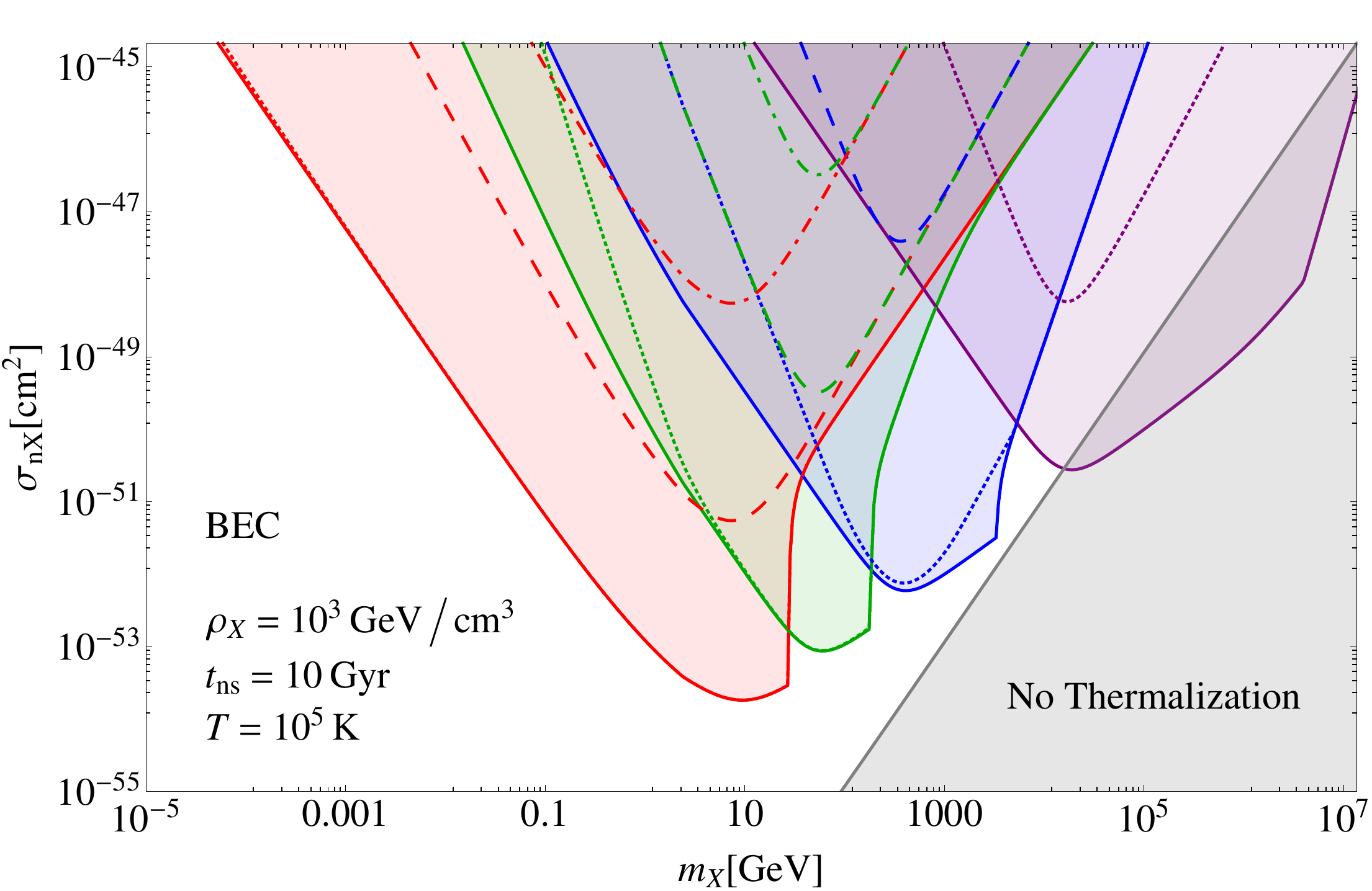}
\caption{Neutron star collapse bounds for annihilating, self-interacting bosonic dark matter that forms a Bose-Einstein condensate at globular
cluster density $\rho_X \sim 10^3 ~ {\rm GeV/cm^3}$. From left to right the red, green, blue, and purple contours denote regions for which the
self interaction parameter $\lambda = \{0,10^{-30},10^{-25},10^{-15}\}$, respectively. Solid, dotted, dashed, and dot-dashed contours denote
annihilation cross sections $\Expect{\sigma_a v}= \{0,10^{-50},10^{-45},10^{-42}\}{\rm cm^3 / s}$, respectively.
}
\label{mvcsx}
\end{figure}

Figure \ref{mvcsx} displays the exclusion contour in the ($m_X$, $\sigma_{nX}$) plane if the dark matter can
form a BEC, assuming that old neutron stars have lifetime $t_{ns} = 10 ~{\rm Gyr}$, core temperature $T={\rm 10^5 ~ K}$, and ambient dark matter
density $\rho_X = 10^3~ {\rm GeV/cm^3}$ (this is an estimate for the dark matter density at the center of globular clusters~\cite{bertone,mccullough}).
The various contours are for different choices of the self-interaction parameter $\lambda$ and the annihilation cross section $\langle \sigma_a v \rangle$
(thermally-averaged at the temperature of the neutron star).
We note that an order of magnitude increase in $\langle \sigma_a v \rangle$ corresponds to an order of magnitude
relaxation of the bound, except in regions
where the bound is significantly affected by dark matter accretion.
Moreover, as the capture rate depends on $\rho_X$ only through the factor $\rho_X \times \sigma_{nX}$, a scaling of the ambient dark matter
density simply rescales the bound on $\sigma_{nX}$.
For example, if the dark matter density at the center of globular clusters is as small as $0.3~{\rm GeV/cm^3}$, then the bound
on $\sigma_{nX}$ would be weakened by a factor of $\sim 3000$.
Note that observations of old neutron stars can only provide
constraints in the region $\sigma_{nX} \leq \sigma_{sat.}$.
For the entire relevant range of masses, bounds from neutron stars become
completely unconstraining if $\langle \sigma_a v \rangle \times f \gtrsim
10^{-38}~{\rm cm^3 /s}$ (thermally averaged at the temperature of the neutron star),
because the neutron star can never capture enough dark matter for black hole collapse to occur.

As previously noted, if the self-interaction term is increased (contours farther right in Figure \ref{mvcsx}), the bound on high mass dark matter improves.
As the Chandrasekhar bound increases with the self-interaction coupling, there is less Hawking radiation at the formation of higher mass black holes, and
the black hole growth condition (eq.~\eqref{thebound}) is met for higher masses and lower scattering cross sections.

Figure~\ref{mvcsx} also shows the excluded region (eq.~\eqref{ttherm}) within which dark matter captured by a neutron star will not thermalize,
\bea
\sigma_{nX} < 1.1 \times 10^{-60} ~{\rm cm^2} \lp \frac{m_X}{\rm GeV}\rp^2 \lp \frac{10^5 \rm K}{T} \rp \lp \frac{\rm 10 ~Gyr}{t_{th}} \rp.
\eea
In the plot we assume a thermalization time scale of $t_{th} \sim$ Gyr.
For $m_X < 28~ {\rm GeV}$, we reproduce the bounds of \cite{mcdermott}.
This analysis shows that old neutron stars in the center of globular clusters with a dark matter density
$\rho_X = 10^3 ~ {\rm GeV/cm^3}$ \cite{bertone,mccullough} provide a bound on non-annihilating bosonic dark matter competitive with
planned terrestrial direct detection experiments for dark matter masses up to $m_X \sim 10^7 ~ {\rm GeV}$ \cite{darkside,xenon1t}. We note that
future detection
of neutron stars in regions of dark matter density larger than $10^3 ~ {\rm GeV/cm^3}$ will result in appropriately rescaled bounds.

\subsection{Dark Matter Annihilation From a BEC State}

Thus far we have modeled dark matter annihilation as arising from a uniform distribution within
the thermalization radius.  However, if dark matter forms a BEC, then a significant fraction of the dark matter
will be in the ground state.  The BEC will be much denser; if dark matter can
annihilate from this state, bounds on dark matter arising from observations of old neutron stars will be
significantly weakened, if not entirely removed.  When the BEC first forms, $r_c / r_{th} \sim 10^{-6}$.
Thus, even a very
small cross-section for dark matter to annihilate from the BEC state can result in a depletion of dark matter large
enough to prevent a black hole from forming.  Moreover, annihilation of dark matter in the BEC can
heat the dark matter, also potentially obstructing black hole formation.

However, one cannot determine the cross section for annihilation from the BEC state from $\langle \sigma_A v \rangle$
in a model-independent way, since $\langle \sigma_A v \rangle$ is determined by thermally-averaging the cross section
at the temperature of the neutron star, $T\sim 10^5~{\rm K}$.  Dark matter in the BEC state is initially much less energetic
than dark matter in the thermalized region; the cross section for annihilation of dark matter in the ground state is
thus model-dependent.  As dark matter continues to accumulate in the BEC state, the dark matter will eventually become self-gravitating.
Once the dark matter is self-gravitating, the size of the BEC will decrease as more particles fall into the ground
state, causing the density of the BEC state to increase and causing the dark matter particles to have larger kinetic
energy.  By the time the Chandrasekhar bound is crossed, the dark matter will be relativistic.
A more complete discussion of dark matter annihilation in the BEC state is beyond the scope
of this work.

\section{Conclusions}

We have studied the constraints that old neutron stars place on bosonic dark matter, allowing for
self-interactions, decay, and self-annihilation of the bosonic dark matter. Observations of old neutron stars imply bosonic
dark matter with a mass $\sim \rm kev - 10^6 ~GeV$ detected at terrestrial experiments will have a minimum
annihilation or self-interaction term. For example, we show that a neutron star of age $t_{ns}=10~{\rm Gyr}$ found
in a globular cluster with dark matter density $\rho_X = 10^3~ {\rm GeV/cm^3}$ will not constrain
bosonic dark matter if the dark matter has an annihilation cross section $\Expect{\sigma_a v} \gtrsim 10^{-38}~ {\rm cm^3/s}$
(thermally-averaged at the temperature of the star, $T \sim 10^5~{\rm K}$).
If dark matter has even a small cross section to annihilate from a BEC state, then constraints
from neutron stars can be weakened even more.
These bounds are thus most relevant if the unbroken symmetry which stabilizes an asymmetric dark matter candidate is
continuous; if it is broken to a $Z_2$ symmetry (even weakly), then self-annihilation is permitted and
these bounds can be weakened considerably.
Conversely we demonstrate that permitted dark matter decay, which is constrained by the evolution of dark halos, will not
significantly relax neutron
star bounds on bosonic dark matter.

We also show that even small self-interaction terms can dramatically weaken bounds on
asymmetric dark matter.  These bounds are thus most constraining if there exists some
(at least approximate) symmetry which can suppress a quartic self-interaction term.  However,
very small self-interactions can result in even more constraining bounds, by causing the
formation of larger black holes which grow rapidly.

It is interesting to note that these bounds can extend up to large $m_X$.  For such large
masses, one cannot easily tie the dark matter asymmetry to the baryon asymmetry.  Nevertheless,
asymmetric dark matter with a small (or vanishing) annihilation cross section provides an
interesting candidate for non-thermal dark matter, and observations of old neutron stars can
provide significant constraints on these models.

Note added: While this paper was being completed,~\cite{Kouvaris:2012dz} appeared, which
also discusses the effect of dark matter accretion on the evolution of a black hole.
In~\cite{Kouvaris:2012dz}, it is also argued that neutron star observations cannot bound dark matter particles
with a large mass.  Note that this argument refers to potential bounds
on dark matter which self-gravitates in the thermalized region before forming a BEC.  The
bounds we have described for BEC dark matter at high mass are not affected by this argument.
These bounds instead arise from a consideration of the effect of self-interactions, and the
effect on the black hole's evolution
of the accretion of dark matter in the BEC phase.

\vskip .1in

{\bf Acknowledgments}

We are grateful to C.~Vause and H.~Yu for useful discussions. J.~K.~thanks the Indian Institute of
Science for its hospitality while this work was being completed. J.~B.~thanks the Institute for
Gravitation and the Cosmos at Penn State for its hospitality.
This work is supported in part by
Department of Energy grant DE-FG02-04ER41291.

\appendix
\label{appendix}

\section{Dark Matter Capture and Annihilation Equilibrium} \label{equilib}
\begin{figure}[!ht]
\centering
\includegraphics[scale=.75]{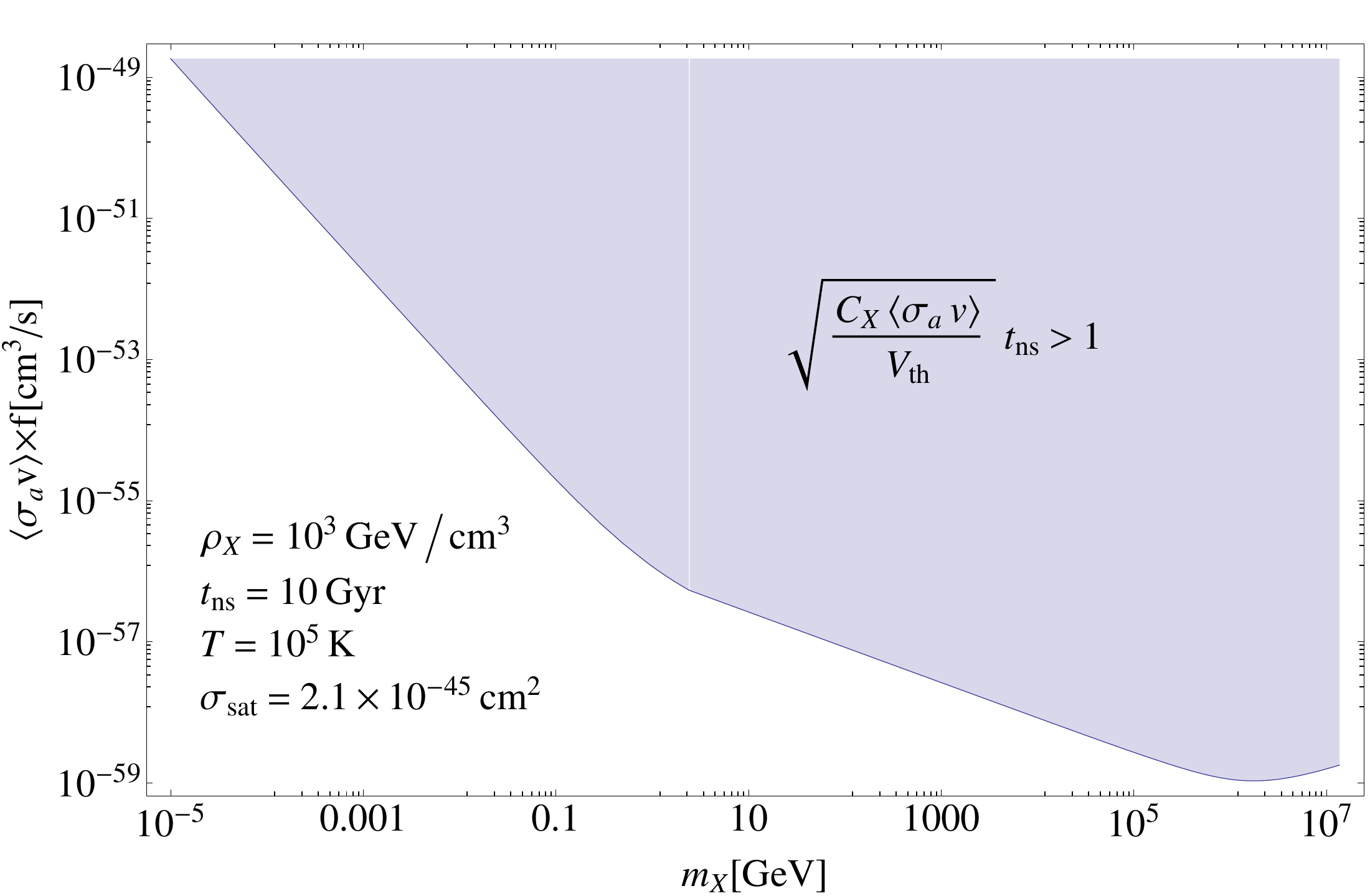}
\caption{In the dark shaded region of the $(m_X, \langle \sigma_a v \rangle \times f)$-plane, the
neutron star is in equilibrium, assuming $t_{ns}=10~{\rm Gyr}$, $T=10^5~{\rm K}$ and $\rho_X = 10^3~{\rm GeV / cm^3}$.
}
\label{sigavmx}
\end{figure}

It is useful to determine the range of parameters for which the total dark matter in the neutron star will reach an equilibrium,
at which point dark matter will annihilate
at the same rate it collects in the neutron star. This happens when the argument of the hyperbolic tangent in eq.~\eqref{nacc} is
greater than unity -- in this range the hyperbolic function will
evaluate to unity and the formula for the number of accumulated dark matter particles simplifies considerably. For $m_X >$ GeV,
\bea
\sqrt{\frac{C_X \langle \sigma_a v\rangle }{ V_{th}}} t_{ns} &\approx& 3.5 \times 10^5 \lp \frac{m_X}{\rm{GeV}} \rp^{1/4}
\lp \frac{\rm 10^5 ~ K}{T} \rp^{3/4}
\lp \frac{t_{ns}}{10~\rm{Gyr}} \rp \beta \nn \\
&\,& \times  \lp \frac{\langle \sigma_a v \rangle}{\rm 10^{-45}~ cm^3/s} \cdot \frac{\rho_X}{{\rm 10^3~ GeV/cm^3}} \cdot
f(\sigma_{nX}) \rp^{1/2} , \label{tanh1}
\eea
and for $m_X <$ GeV,
\bea
\sqrt{\frac{C_X \langle \sigma_a v \rangle }{ V_{th}}} t_{ns} &\approx& 5.1 \times 10^5 \lp \frac{m_X}{\rm{GeV}} \rp^{3/4}
\lp \frac{\rm 10^5 ~ K}{T} \rp^{3/4}
\lp \frac{t_{ns}}{10~\rm{Gyr}} \rp \beta \nn \\
&\, &\times \lp \frac{\langle \sigma_a v \rangle}{\rm 10^{-45}~ cm^3/s} \cdot \frac{\rho_X}{{\rm 10^3~ GeV/cm^3}} \cdot
f(\sigma_{nX}) \rp^{1/2}. \label{tanh2}
\eea
In figure~\ref{sigavmx}, we plot the region in the $(m_X, \langle \sigma_a v\rangle \times f)$-plane such
that the neutron star is in equilibrium, assuming $t_{ns}=10~{\rm Gyr}$, $T=10^5~{\rm K}$ and $\rho_X = 10^3~{\rm GeV / cm^3}$.
Even for small values of $\sigma_{nX}$ and $\langle \sigma_a v \rangle$ (including any values which could be detected by
current and planned observations), we may use the approximation
\bea
N_{acc} &\sim& \sqrt{C_X V_{th}/ \langle \sigma_a v \rangle}
\nonumber\\
&\sim& \sqrt{{4\pi C_X (240~{\rm cm})^3 \over  3 \langle \sigma_a v \rangle}}\left({T \over 10^5 K} {\gev \over m_X} \right)^{3/4} .
\eea
But if dark matter has not reached equilibrium, we instead find
\bea
N_{acc} &\sim& C_X t_{ns} .
\eea

\section{Thermalization Time} \label{thermapp}
For the dark matter particles to achieve self-gravitation, the dark matter must
thermalize with the neutron star on a time scale comparable to the neutron star lifetime. For $m_X \gtrsim {\rm GeV}$,
the thermalization time is \cite{mcdermott}
\beq
t_{th} = 5.4 \times 10^{-6} {\rm years} \lp \frac{m_X}{{\rm GeV}} \rp^2
\lp \frac{10^5~ {\rm K}}{T} \rp f^{-1}, \label{ttherm}
\eeq
where $f= \sigma_{nX} / \sigma_{sat.}$ if $\sigma_{nX} < \sigma_{sat.}$, and $f=1$ otherwise.
Here, $\sigma_{nX}$ is the cross section for DM-neutron interactions,
$\sigma_{sat.} \sim 2.1 \times 10^{-45}~ {\rm cm^2}$ and $T$ is the core temperature of the neutron star.
Thus eq.~\eqref{ttherm} presents a firm exception to the application of the neutron star collapse bound (eq.~\eqref{thebound}): if the dark matter
does not thermalize over the lifetime of the star, it will not form a black hole.

It is important that the saturation of the likelihood of dark matter scattering implies a maximum dark matter mass
for which neutron star bounds are applicable; beyond this
mass the dark matter will not thermalize:
\beq
m_{X}^{(max)} = 1.4 \times 10^7 { \rm GeV} \lp \frac{T}{10^5 ~ \rm{K}} \rp^{1/2}
\lp \frac{t_{ns}}{\rm{Gyr}} \rp^{1/2} \label{mxmax}.
\eeq

\section{Dark Matter Which Does Not Form a BEC} \label{AppendixNonBEC}

\begin{figure}[ht]
\centering
\includegraphics[scale=.9]{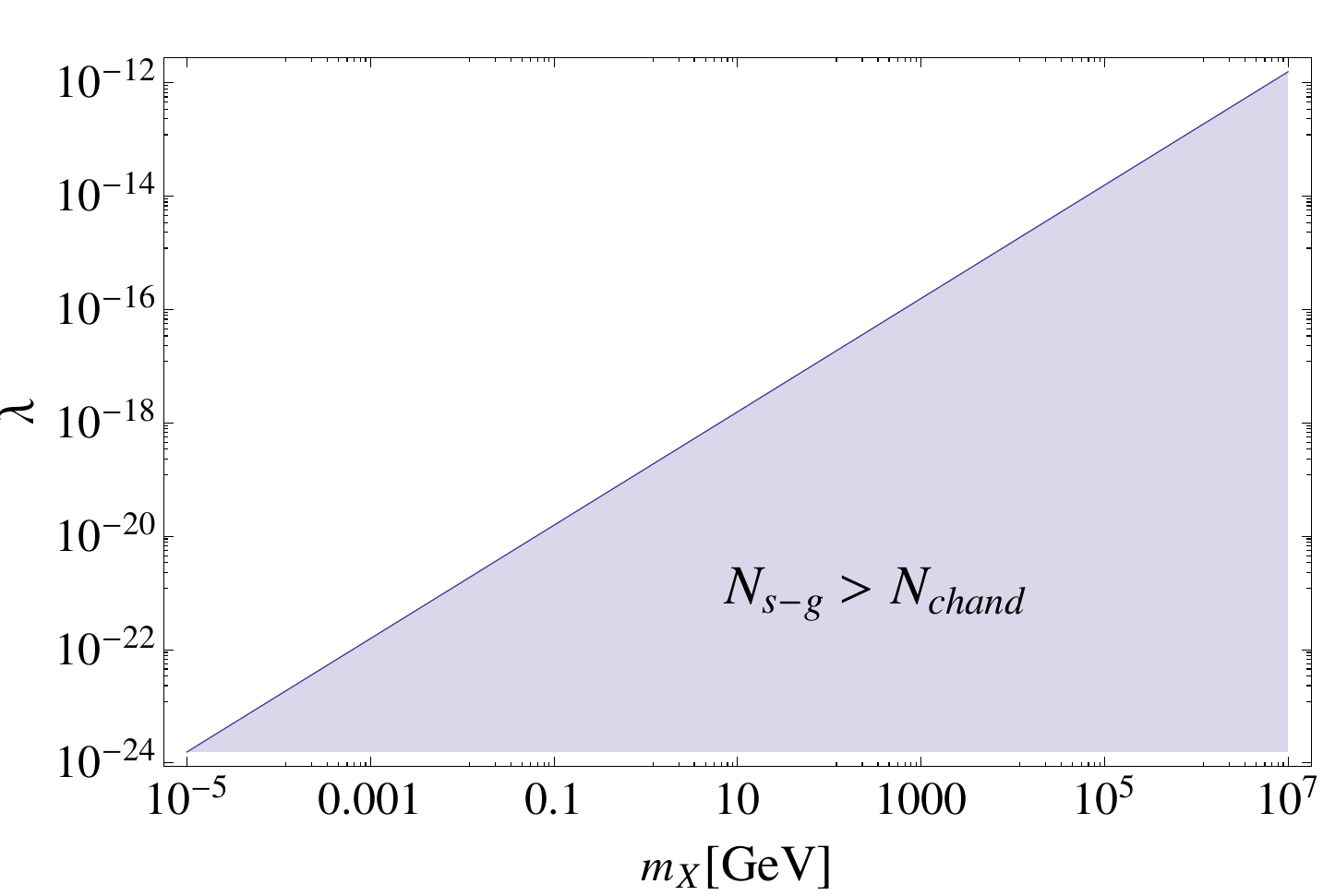}
\caption{Parameter space in the $(m_X, \lambda)$ plane where $N_{s-g}^{(th.)} > N_{chand}$, if a BEC does not form.
}
\label{mxvlam}
\end{figure}

If dark matter does not form a BEC, then the analysis is slightly different.
The number of dark matter particles in the thermalized region
required to achieve self-gravitation is given by
\beq
N_{s-g}^{(th)} \simeq \frac{4 \pi r_{th}^3}{3 m_X}  \rho_b \simeq 4.8 \times 10^{46} \lp \frac{T}{10^5~ \rm{K}} \rp^{3/2}
\lp \frac{\rm{GeV}}{m_X} \rp^{5/2}. \label{nsg-nonBEC}
\eeq
For most regions of interest, dark matter which does not form a BEC will collapse when it becomes self-gravitating.
But if $\lambda$ is large enough, thermalized boson distributions will collapse only when they reach the
Chandrasekhar limit. In figure \ref{mxvlam}, we plot the region where $N_{s-g}^{(th.)} > N_{chand}$ in the
$(m_X, \lambda)$-plane for $T=10^5~{\rm K}$, in the case where a BEC does not form.

The number of particles required for a black hole to form in the case where a BEC does
not form is then given by:
\bea
N_{BHforms}^{(th)}(m_X, \lambda, T) &=& \max [N_{chand} , N_{s-g}^{(th)}].
\eea
If $N_{chand} > N_{s-g}$, then the initial mass of the black hole is the same as the value given
in eq.~\ref{eq_MBHI_BEC}.  But if $N_{s-g} > N_{chand}$, then the initial black hole mass is given by
\bea
M_{bhi}^{(therm)} &=&m_X N_{s-g}^{(therm)} = 4.8 \times 10^{46}~\gev \lp \frac {T}{10^5 {~ \rm K}}\rp^{3/2}
\lp \frac{ \rm GeV}{m_X} \rp^{3/2} .
\eea
In this case, if the black hole grows, the time required for the neutron star to be destroyed is
\bea
t_{nscollapse}^{(therm)} &=& 5.1 \times 10^{-4} ~{\rm years} \lp \frac{m_X}{{\rm GeV}} \rp^{3/2} \lp \frac{10^5 ~ \rm K}{T} \rp^{3/2},
\eea
and the black hole will quickly destroy the neutron star.
If the black hole evaporates, the time required to complete the evaporation process is given by
\bea
t_{evap}^{(s-g)} &=& 5120 \pi G^2 \left(4.8 \times 10^{46}  \lp \frac{ \rm GeV^{5/2}}{m_X^{3/2}} \rp\right)^3
\lp \frac {T}{10^5 {~ \rm K}}\rp^{9\over 2}
\nonumber\\
&=& 1.6\times 10^{27} {~\rm Gyr} \lp \frac {T}{10^5 {~ \rm K}} {\gev \over m_X}\rp^{9\over 2}.
\eea
If dark matter does not form a BEC, then the black hole will evaporation time will be
short if $m_X \gg 10^4~\gev$.

The dark matter thermalization radius is much larger than the black hole impact parameter; thus dark matter will not be efficiently
captured by a black hole for dark matter that does not form a BEC.  Instead, the black hole will continue to capture dark matter through
Bondi accretion.
The dark matter accretion rate can then be written in terms of the coupled differential equations.
\bea
\lp {dM_{bh}\over dt} \rp_{DM} &=&  \frac{3 m_X N_{DM} (G M_{bh})^2}{v_s^3 r_{th}^3}
\nonumber\\
\lp {dN_{DM} \over dt} \rp &=& C_X  - {1\over m_X}\lp {dM_{bh} \over dt} \rp_{DM}.
\eea

However in the case of non-BEC dark matter black holes, the Hawking radiation rate exceeds the baryonic Bondi accretion rate only for
\beq
m_X > 5.8 \times 10^6 ~{\rm GeV} \lp \frac{T}{10^5 ~\rm{K}} \rp,
\eeq
To accrete enough dark matter for the dark matter Bondi accretion rate to be of the same order of magnitude as the
baryonic Bondi accretion rate would require about as much time as was required for the black hole to form in the
first place.
But as we have seen, for the range of masses where the Hawking radiation rate dominates when the black hole
is formed, the black hole will evaporate relatively quickly, before the dark matter accretion rate becomes appreciable.
\begin{figure}[!!ht]
\centering
\includegraphics[scale=.75]{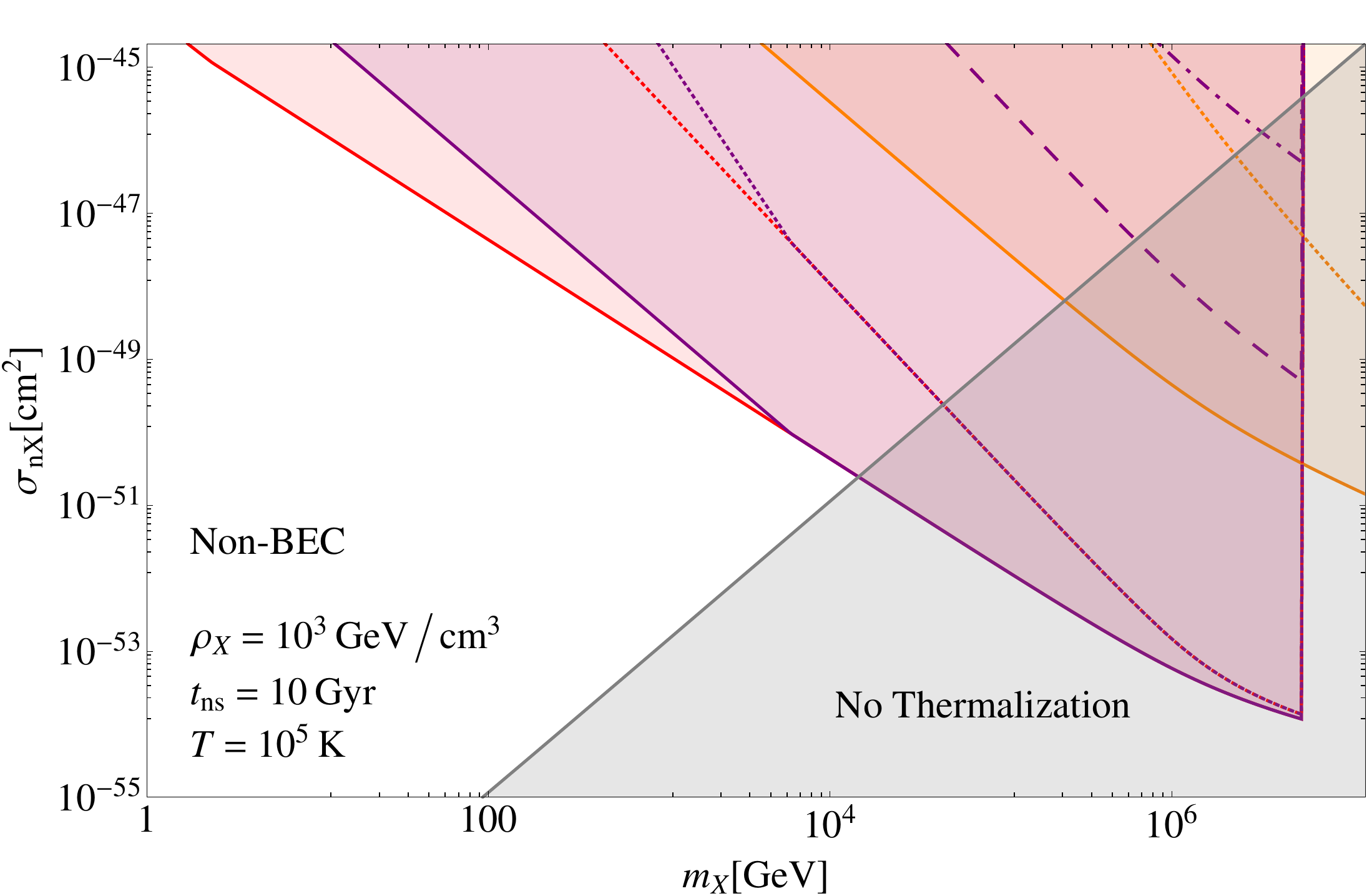}
\caption{Neutron star collapse bounds for annihilating, self-interacting bosonic dark matter that does not form a Bose-Einstein condensate.
From left to right the red, purple, and orange contours denote regions for which the self interaction parameter $\lambda = \{0,10^{-15},10^{-5}\}$,
respectively. Solid, dotted, dashed, and dot-dashed contours denote annihilation cross sections
$\Expect{\sigma_a v}= \{0,10^{-50},10^{-45},10^{-42}\}{\rm cm^3 / s}$, respectively.}
\label{mvcsxnobec}
\end{figure}
We thus
find that, for the case where dark matter does not form a BEC, we may essentially ignore the effect of dark matter
accretion on the evolution of the black hole and the condition for a black hole to destroy the neutron star becomes,
\bea
0 < \left. {dM_{BH} \over dt}\right|_{M_{BH}=M_{BHi}} &\simeq&  \frac{4 \pi \rho_b (G m_X {\rm Max}[N_{chand},N_{s-g}^{(th)}])^2}{v_s^3} \nonumber \\
&\,&- \frac{1}{15360 \pi (G m_X {\rm Max}[N_{chand},N_{s-g}^{(th)}])^2}.
\eea

In Figure \ref{mvcsxnobec}, we plot exclusion contours in the ($m_X$, $\sigma_{nX}$) plane if the dark matter cannot
form a BEC, assuming that old neutron stars have lifetime $t_{ns} = 10 ~{\rm Gyr}$ and core temperature $T={\rm 10^5 ~ K}$,
and assuming an ambient dark matter density of $\rho_X =10^3~{\rm GeV/cm^3}$.
As expected from Figure \ref{mxvlam}, the self-interaction coupling does not affect the bounds substantially until $\lambda \gtrsim 10^{-15}$
-- at this value the number of particles required for self-gravitation is exceeded by the number of particles required by the Chandrasekhar bound.



\end{document}